\documentclass[twocolumn]{aastex631}

\usepackage{amsmath}

\usepackage[T1]{fontenc}
\usepackage[utf8x]{inputenc}

\newcommand{\ratio}{ D_\mathrm{L, GW}/D_\mathrm{L, EM} }
\newcommand{\ratioeq}{ \frac{D_\mathrm{L, GW} }{D_\mathrm{L, EM} } }

\newcommand{\ratiobias}{ b_\mathrm{GW}/b_\mathrm{gal} }

\newcommand{\dlgw}{ D_\mathrm{L, GW} }
\newcommand{\dlem}{ D_\mathrm{L, EM} }

\newcommand{\bgw}{ b_\mathrm{GW} }
\newcommand{\bg}{ b_\mathrm{gal} }

\newcommand{\ngw}{ n_\mathrm{GW} }

\begin{document}

\title{Learning how to surf: \\ Reconstructing the propagation and origin of gravitational waves with Gaussian Processes}

\author[0000-0003-2796-2149]{Guadalupe Ca\~nas-Herrera}
\email{canasherrera@lorentz.leidenuniv.nl}
\affiliation{%
Leiden Observatory, Leiden University, PO Box 9506, Leiden 2300 RA, The Netherlands}
\affiliation{%
Lorentz Institute for Theoretical Physics, Leiden University, PO Box 9506, Leiden 2300 RA, The Netherlands
}%
\author[0000-0001-7699-6834]{Omar Contigiani}
\email{contigiani@lorentz.leidenuniv.nl}
\affiliation{%
Leiden Observatory, Leiden University, PO Box 9506, Leiden 2300 RA, The Netherlands}
\affiliation{%
Lorentz Institute for Theoretical Physics, Leiden University, PO Box 9506, Leiden 2300 RA, The Netherlands
}%
\author[0000-0002-8496-5859]{Valeri Vardanyan}
\email{valeri.vardanyan@ipmu.jp}
\affiliation{Kavli Institute for the Physics and Mathematics of the Universe (WPI), UTIAS, The University of Tokyo, Chiba 277-8583, Japan}

\begin{abstract}

Soon, the combination of electromagnetic and gravitational signals will open the door to a new era of gravitational-wave (GW) cosmology. It will allow us to test the propagation of tensor perturbations across cosmic time and study the distribution of their sources over large scales. In this work, we show how machine learning techniques can be used to reconstruct new physics by leveraging the spatial correlation between GW mergers and galaxies. We explore the possibility of jointly reconstructing the modified GW propagation law and the linear bias of GW sources, as well as breaking the slight degeneracy between them by combining multiple techniques. We show predictions roughly based on a network of Einstein Telescopes combined with a high-redshift galaxy survey ($z\lesssim3$). Moreover, we investigate how these results can be re-scaled to other instrumental configurations. In the long run, we find that obtaining accurate and precise luminosity distance measurements (extracted directly from the individual GW signals) will be the most important factor to consider when maximizing constraining power. 

\end{abstract}


\section{Introduction} \label{sec:intro}

The first direct detection of gravitational waves (GWs) by \cite{Abbott:2016blz} triggered a rapidly increasing interest in exploiting this new field for cosmological information. GWs alone are not particularly useful because the data provides only a sky position and a measure of the luminosity distance to the source. However, GW sources can serve as powerful cosmological probes when combined with electromagnetic (EM) data, from which redshifts can be extracted. This idea dates back to \cite{Schutz:1986gp} and has two main variations.

The first, simplest, possibility is the observation of so-called \textit{standard sirens} \citep{Holz:2005df}, GW sources with EM counterparts from which a redshift can be observed. As an example, a population of binary neutron stars, such as the already observed GW170817 \citep{TheLIGOScientific:2017qsa}, can be used to reconstruct the luminosity-redshift function and constrain cosmological observables. In this context, the aforementioned observation has already lead to promising results in constraining the Hubble parameter $H_0 = 100h~\mathrm{km/s/Mpc}$ \citep{Abbott:2017xzu}.

The second possibility is to use \textit{dark sirens} with a statistical counterpart. The likely counterpart is identified using the rough sky-localization offered by current GW detectors in combination with deep galaxy catalogs. This technique, sometimes combined with the first, has also been used to extract a measurement of $H_0$ \citep{Soares-Santos:2019irc,Abbott:2019yzh} and a more complex variation of it is the focus of this work. Assuming that both galaxies and the hosts of GW mergers are biased tracers of the same cosmological structure, it is possible to measure a non-zero cross-correlation signal between the two (see \autoref{sec:Cells}). At the linear level, in particular, the description of this signal is especially simple. It should be noted that this idea has been explored by others before, sometimes using a different formalism \citep[see, e.g., ][]{Mukherjee:2018ebj, Scelfo:2018sny} and that similar methods have already been employed to provide a measurement of $H_0$ \citep{Finke:2021aom}. 

We also mention that, in principle, the same idea can be used in the absence of resolved events. Previous works, however, have shown that a detection of this effect from a stochastic background signal is unlikely to happen soon due to the low signal-to-noise ratio (SNR) attainable by current and proposed experiments covering the optimal wavelength range \citep{Canas-Herrera:2019npr, Alonso:2020mva}.

In this paper, we discuss the possibility of using the non-zero correlation between the distribution of EM galaxies and resolved GW mergers to jointly extract information about the two main quantities in the field of GW cosmology: the luminosity distance as a function of redshift, describing the propagation of gravitational waves across cosmic time, and their linear bias, describing their clustering properties.  

It is already established that GWs carry the potential of constraining the fundamental laws of gravity. This is the case because the propagation of gravitational waves in modified gravity scenarios differs from predictions of the general theory of relativity (GR) in multiple ways \citep[see, e.g., ][]{Deffayet:2007kf, Garoffolo:2019mna, Ezquiaga:2020dao}. One of the clear signatures is the speed of tensor modes which can be both sub- and superluminal as opposed to the GR case, where GWs propagate at the speed of light. The tight constraints on speed deviations imposed by the multimessenger observations of GW170817, for example, have ruled out a wide parameter space of otherwise viable scalar-tensor theories of gravity \citep{Lombriser_2016,Sakstein_2017,Ezquiaga:2017ekz, Creminelli:2017sry,Baker_2017}. Similarly, implications for bi-gravity models have also been demonstrated in the literature \citep[see][]{Max_2017,Max_2018,Akrami:2018yjz,Belgacem:2019pkk}.

Another striking difference between modified gravity and GR is the modified friction of GWs \citep[][]{Amendola_2018,Belgacem:2018lbp}. This feature arises in models with non-minimal coupling of a scalar field and gravity which manifests itself as a redshift-dependent gravitational coupling. As a result, the inferred luminosity distance to GW sources differs from the corresponding EM luminosity distance. This interesting phenomenon is explained in \autoref{sec:GWs}, and in this work, we will investigate the possibility of testing this hypothesis. 

The discovery of the first LIGO-Virgo binary black hole has been used to motivate alternative scenarios where the binary did not represent the endpoint of stellar evolution, but originated either as a pair of primordial black holes (PBH) or some exotic compact objects \citep[see, e.g.,][]{Bird:2016dcv,Sasaki:2016jop, CalderonBustillo:2020srq}. In particular, the last half-decade has seen a resurgence in interest for PBHs \citep[see, e.g.,][]{Clesse:2016vqa, Sasaki_2018,Raccanelli:2016cud, Raccanelli:2017xee}. The main difference between the PBH and stellar evolution scenarios is the spatial distribution of GW mergers, measurable both in the redshift and sky distribution of the sources. In \autoref{sec:gwsources} we expand on how to model this difference through the linear bias and present a few models used in this work.

In addition to our formalism, for both the clustering and modified gravity effects we investigate a possible method to precisely reconstruct the redshift evolution of these quantities in the upcoming decades. Our study is based on Gaussian processes (GPs), a well-known hyper-parametric regression method \citep{ML}. The structure and implementation of this pipeline are presented in \autoref{sec:reconstruction}. 

GPs have been widely used in the literature to reconstruct the shapes of physical functions such as the dark energy equation of state $w(z)$ \citep{Shafieloo_2012,Gerardi:2019obr}, the primordial inflaton's speed of sound \citep{Canas-Herrera:2020mme} or the mass function of the merging binary black hole systems \citep{Li:2021ukd}. GPs are very useful for such functional reconstructions due to their flexibility and simplicity. In general, binned reconstructions, such as the framework employed by \cite{Crittenden_2009,Crittenden_2012} and \cite{Zhao_2012}, as well as parametric reconstructions with high-degree polynomial functions can be reproduced using GPs with a handful of hyper-parameters.

Unless stated otherwise, our fiducial cosmology is based on the best fit results from Planck 2018 \citep{Akrami:2018vks}. In our analysis, we use \texttt{COLOSSUS} \citep{Diemer:2017bwl} and \texttt{Astropy} \citep{Robitaille:2013mpa, Price-Whelan:2018hus} for cosmological calculations, \texttt{sklearn} \citep{sklearn} for the GP implementation, \texttt{emcee} \citep{ForemanMackey:2012ig} as our posterior sampler and \texttt{GetDist} \citep{getdist} to plot the final contours. Our analysis pipeline is made publicly available\footnote{\url{https://github.com/valerivardanyan/GW-Cosmo}}. 

The paper is organized as follows. In \autoref{sec:GWs} we summarize the essential concepts concerning the GW propagation in modified gravity models. In \autoref{sec:Cells} we detail the modelling of the clustering correlations between the GW source population and galaxies. In \autoref{sec:reconstruction} we explain our reconstruction pipeline for the GW luminosity distance and the bias. Our findings are presented in \autoref{sec:results} and further discussed in \autoref{sec:discussion}.

\section{Gravitational wave propagation}
\label{sec:GWs}
In GR, and around a background Friedmann-Lemaitre-Robertson-Walker (FLRW) metric, the amplitude of GWs evolves according to
\begin{equation}
\label{eq:GR-propagation}
    h^{\prime\prime}_\alpha + 2\mathcal{H} h^\prime_\alpha - \vec{\nabla}^2 h_\alpha = 0,
\end{equation}
where $h_\alpha$ denotes the amplitude of either polarization ($\alpha \in [\times, +]$), primes denote derivatives with respect to the conformal time, and $\mathcal{H}$ is the conformal Hubble function. In this equation, the prefactor of the gradient term controls the propagation speed, which we have set to coincide with the speed of light in $c=1$ units.

The second term is the standard cosmic friction term which causes the strain amplitude to decay as ${h_\alpha(z) \propto D_\mathrm{L}^{-1}(z)}$, with $D_L$ being the FLRW luminosity distance:
\begin{equation}
    D_\mathrm{L}(z) = (1+z) \int_0^z \; \frac{\mathrm{d}\tilde{z}}{H(\tilde{z})},
    \label{eq:dl}
\end{equation}
where the Hubble function $H(z)$ is given in terms of the Hubble constant $H_0$, present-day dark matter abundance $\Omega_\mathrm{m}$ and dark energy abundance $\Omega_\mathrm{DE}(z)$ as
\begin{equation}
\label{eq:hubble}
    H(z) = H_0 \left[ \Omega_m (1+z)^3 + \Omega_{DE}(z)  \right].
\end{equation}
Throughout this paper we assume a constant equation of state $w_0$ for dark energy, such that its energy density is given by
\begin{equation}
\label{eq:de}
    \Omega_{DE}(z) = (1 - \Omega_\mathrm{m}) (1 + z)^{3(1 + w_0)}.
\end{equation}
The standard $\Lambda CDM$ cosmology corresponds to ${w_0 = -1}$.

It is now established that modifications of GR can affect the propagation of GWs. The important effect for us is the modified friction term with respect to the GR expectation in Equation~\eqref{eq:GR-propagation},
\begin{equation}\label{eq:mod-propagation}
    h^{\prime\prime}_\alpha + \left[2 + \alpha_M(z)\right]\mathcal{H} h^\prime_\alpha - \vec{\nabla}^2 h_\alpha = 0,
\end{equation}
where we have again imposed the GW speed to be unity as suggested by observations. The modified friction term introduces a new scaling $h_\alpha(z) \propto 1/\dlgw(z)$, with $\dlgw(z) \neq D_\mathrm{L}(z)$ for non-zero $\alpha_M(z)$. The luminosity distance to GW events can be written as:
\begin{equation}
    \ratioeq\left(z\right) = \exp \left\{ - \frac{1}{2}\int_0^z\mathrm{d}\tilde{z} \; \frac{\alpha_M(\tilde{z})}{(1 + \tilde{z})} \; \right \}.
    \label{eq:dlratio}
\end{equation}
In this work, we assume that the luminosity distance for EM sources $\dlem$ is unaffected and is equal to the expression in Equation~\eqref{eq:dl}.
The function $\alpha_M$ corresponds to the running of the effective Planck mass, i.e.,
\begin{equation}
    \alpha_M = \frac{d \log (M_\mathrm{eff}/M_\mathrm{P})^2}{d \log a},
\end{equation}
where $M_\mathrm{P}$ is the Planck mass and $M_\mathrm{eff}$ is its effective value at redshift $z = 1/a-1$. This function encodes information about extensions of GR such as scalar-tensor theories \citep{Horndeski:1974wa, Bellini:2014fua} or, more broadly, quantum gravity \citep{Calcagni:2019ngc}. The modified friction term is also a natural prediction of non-local modifications of gravity \citep[][]{Dirian_2016}

From an effective field theory point of view $\alpha_M(z)$ is a free function of order unity. In practical studies of modified gravity and dark energy, however, $\alpha_M$ is often assumed to take simple parametric forms. The main guiding principle is the assumption that its effects should be negligible in the early universe, which prompts to choose $\alpha_M(z)$ to be proportional either to the dark energy abundance or simply to some power of the scale factor $a$.    

Such parametrizations make it possible to find a closed form expression for the ratio in Equation~\eqref{eq:dlratio} and have inspired a widely used parametrization of the ratio as a monotonic deviation which goes to $1$ at present day \citep{Belgacem:2018lbp}
\begin{equation}
    \ratioeq(z) = \Xi_0 + \frac{1-\Xi_0}{(1+z)^n}.
    \label{eq:dlratioxi}
\end{equation}
In this expression, $\Xi_0$ and $n$ are two constant parameters, which are typically of order $\mathcal{O}(1)$.

\section{Angular power-spectra}
\label{sec:Cells}
\subsection{GW sources}
\label{sec:gwsources}

We consider GW mergers with a distribution in redshift written as
\begin{equation}
\label{eq:ngw}
    \ngw(z) = \frac{n_0}{1+z},
\end{equation}
where $n_0$ corresponds to the comoving number density of observed events as a function of redshift, and the term $(1+z)$ takes into account the cosmological time dilation. In our analysis, we use a constant value of $n_0 \approx 3\times10^{-6}$ $h^3$Mpc$^{-3}$ (with $h$ denoting the usual normalized Hubble constant), motivated by current LIGO constraints \citep{Abbott:2020gyp}. 

For a given selection of sources along the line of sight, the average number of projected sources can be written using the comoving distance $\chi(z)$: 
\begin{equation}
    \bar{n}_\mathrm{gw} = \int_0^\infty dz \frac{\chi^2(z)}{H(z)}S(z) \ngw(z).
    \label{eq:avgproj}
\end{equation}

The function $S$ encodes the selection and the scatter due to observational errors. In this paper, simple bins in a range $[D_\mathrm{L, min},D_\mathrm{L, max}]$ are used and we assume a lognormal distribution with fixed scatter $\sigma_{\ln D}$ for the individual sources \citep{Oguri:2016dgk}. In this case, $S$ can be written as:
\begin{equation}
    S(z) = \frac{1}{2} \left[ x_\mathrm{min} (z) - x_\mathrm{max}(z) \right],
\end{equation}
with
\begin{equation}
    x_\mathrm{min}(z) = \mathrm{erfc} \left[ \frac{\ln D_\mathrm{L, min}  - \ln \dlgw(z)}{\sqrt{2}\sigma_{\ln D}} \right],   
\end{equation}
and similarly for $x_\mathrm{max}$. Including this effect makes $S$ resemble a top-hat function with damping tails dictated by $\sigma_{\ln D}$.

The angular power spectrum of these sources can be written using the Limber approximation
\begin{equation}
\begin{split}
    C_\mathrm{GW}(\ell) = &\int_0^\infty dz \; \frac{H(z) }{\chi^2(z)} W_\mathrm{GW}^2(z) \\ &\bgw^2(z) P\left(\frac{\ell+1/2}{\chi(z)}, z\right),
    \label{eq:autogw}
\end{split}
\end{equation}
where $P(k, z)$ is the matter power-spectrum at redshift $z$ and comoving scale $k$, $\bgw$ is the bias of the GW sources, and the window function can be written as
\begin{equation}
    W_\mathrm{GW} (z) = \frac{\chi^2(z)}{H(z)} \frac{\ngw (z)}{\bar{n}_\mathrm{GW}} S(z).
\end{equation}

For the purpose of illustration, we will make use of a few simple parametrization for the GW bias. We will consider either a constant bias $\bgw$ with a value of order $\mathcal{O}(1)$ or a more complex form:
\begin{equation}
    \label{eq:biasgw}
    \bgw(z) = b_0\left(1 + \frac{1}{D(z)}\right),
\end{equation}
where $D(z)$ represents the growth factor. The first model, with its low constant value, mimics a PBH origin for the mergers  \citep{Bird:2016dcv, Raccanelli:2016cud}, while the second mimics the stellar evolution case by tracking the galaxy linear bias \citep{Oguri:2016dgk}. 

\subsection{Galaxies}

Similarly to the GW population, we again assume a constant comoving number density of galaxies. Throughout our analysis we fix
\begin{equation}
    n_\mathrm{gal} (z) = 10^{-3}h^3\mathrm{Mpc}^{-3},
\end{equation}
and we write the autocorrelation signal of galaxies under the Limber approximation as
\begin{equation}
\begin{split}
    C_\mathrm{gal}(\ell) = &\int_0^\infty dz \; \frac{H(z) }{\chi(z)^2} W_\mathrm{gal}^2(z) \\ &\bg^2(z) P\left(\frac{\ell+1/2}{\chi(z)}, z\right).
    \label{eq:autog}
\end{split}
\end{equation}
In this expression the definition of $W_\mathrm{gal}$ is the same as $W_\mathrm{GW}$ used in the previous section except for using $n_\mathrm{gal}(z)$, a different selection function, and $\bg(z)$ is the linear galaxy bias. In our analyses, we assume a known galaxy bias in the form of
\begin{equation}
\label{eq:bg}
    \bg(z) = 1 + \frac{1}{D(z)}.
\end{equation}
In general, this function is expected to be accurately measured from the galaxy autocorrelation signal alone.

In this paper, we employ a top-hat selection function for $W_\mathrm{gal}$, which assumes no uncertainty in galaxy redshift estimates. This choice mimics a spectroscopic galaxy survey or a general redshift survey with negligible uncertainties. As an example, another choice commonly found in the literature is a Gaussian distribution $\mathcal{N}(z, \sigma_\mathrm{gal})$, where $\sigma_\mathrm{gal}$ should be much larger than the expected redshift uncertainty for each individual galaxy.

By combining the distribution of GW sources and galaxies one can construct a cross-correlation map. In our formalism, we write the cross-correlation between a GW bin $i$ and a galaxy bin $j$ (fully specified by their respective window functions) as:
\begin{equation}
\begin{split}
    C_\times^{ij}(\ell) &= \int_0^\infty \mathrm{d}z\frac{H(z) }{\chi^2(z)}W^i_\mathrm{GW}(z)W^j_\mathrm{gal}(z) \\
    &\times \bgw (z) \bg (z) P\left(\frac{\ell+1/2}{\chi(z)}, z\right).
    \label{eq:cc}
\end{split}
\end{equation}

We conclude this section by pointing out that the power spectra in Equations~\eqref{eq:autog}, \eqref{eq:autogw} and \eqref{eq:cc} do not include relativistic terms and do not capture the effects of evolution and lensing bias \citep[see, e.g., ][for a detailed treatment]{Scelfo:2018sny, Scelfo:2020jyw}. Specifically, while the effects of lensing are expected to be negligible at the redshifts considered here \citep[z<3, see, e.g., ][]{Oguri:2016dgk, Contigiani:2020yyc}, the same is not true for relativistic effects. Therefore, we choose not to consider small values of $\ell$ in our analysis since the signal at these large angular scales is dominated by them.

\section{Reconstructing GW physics}
\label{sec:reconstruction}

The primary goal of the paper is to demonstrate how to reconstruct the properties of GW propagation and source clustering as a function of redshift. We do so by showing how to recover an assumed fiducial model by using mock angular power spectra with cosmic-variance or shot-noise limited uncertainties.

\begin{figure}[htbp]
    \centering
    \includegraphics[width=\columnwidth]{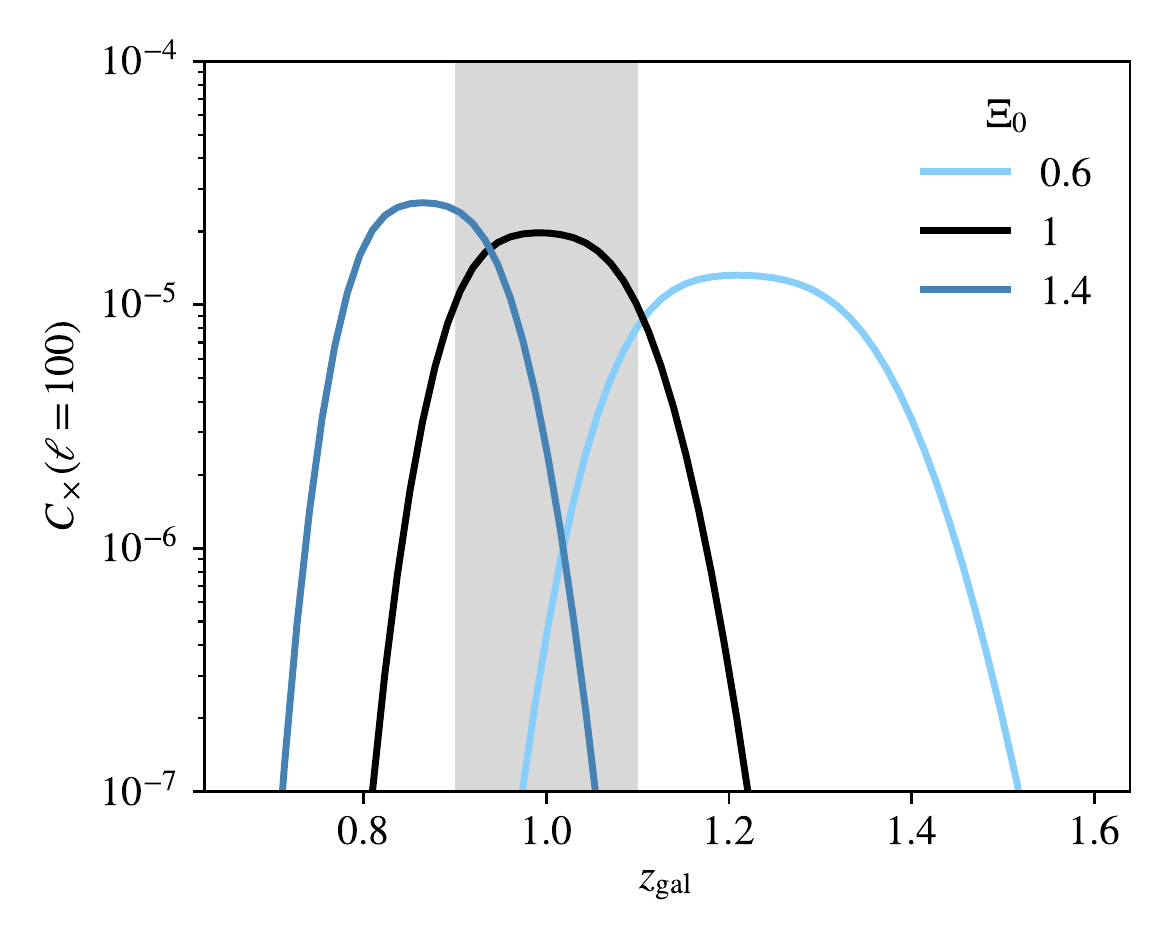}
    \caption{The cross-correlation signal between GW sources at $z=[0.9, 1.1]$ (shaded area) and galaxies at different redshifts ($z_\mathrm{gal}$). If the luminosity distance ratio $\dlgw/\dlem(z)$ in Equation~\eqref{eq:dlratioxi} is different from its GR assumption ($\Xi_0\neq 1$), the location of the predicted cross-correlation peak is also affected.}
    \label{fig:cross_corr_demo}
\end{figure}

Our methodology hinges on the fact that by cross-correlating a GW luminosity distance bin with multiple galaxy redshift bins we can determine the redshift of the GW sources by matching the clustering properties of the two at the true redshift \citep{Oguri:2016dgk, Bera:2020jhx}.

We demonstrate this idea in \autoref{fig:cross_corr_demo}, where we have considered GW sources located at redshift $[0.9, 1.1]$ in a GR cosmology where $\dlgw(z) = \dlem(z)$. In this figure, we show the expected cross-correlation signal between the angular distribution of these sources and the angular distribution of galaxies located at various redshifts. As expected, in GR ($\Xi_0 = 1$) the signal peaks inside the correct redshift range (shaded area). However, as we depart from the GW luminosity distance relation, the location of this peak is affected.

\subsection{Mock data and fiducial model}

In this section, we describe the recipe used to generate the mock angular power spectra ($C_\mathrm{gal}, C_\mathrm{GW}$, and $C_\times$) that are fed into our reconstruction pipeline together with their error covariance matrix. When describing real data, these angular power-spectra are extracted from the autocorrelation and cross-correlation maps representing the sky distribution of galaxies and GW sources. The recipe has three main ingredients: the details of the fiducial model, a description of the instrumental configuration and a definition of the dominant source of error.

The first ingredient is the fiducial model. Our decision in this case is based on the results of \cite{Baker:2020apq}, where present-day constraints on the function $\alpha_M$ appearing in Equation~\eqref{eq:dlratio} are presented. As shown in \cite{Belgacem:2019pkk}, the results of the $\alpha_M\propto a$ parametrization found in that work can be mapped to the $\Xi(z)$ function in Equation~\eqref{eq:dlratioxi}. Using this transformation, we find that the $3\sigma$ upper limit roughly corresponds to
\begin{equation}
        \Xi_0 \lesssim 1.4,
\end{equation}
with $n = 1$. Thus, we assume a fiducial model with $\Xi^\mathrm{fid}_0 = 1.4$ and $n^\mathrm{fid} = 1$, representing the limit of our present understanding.

The second ingredient of our forecast is the instrumental configurations. The size of our data vector is given by the number of multipoles $\ell$ and window functions that we include in our analysis. Since both are largely dictated by observational considerations, in this work we assume an optimistic combination of a network of three Einstein Telescopes \citep{Maggiore:2019uih, 2019CQGra..36v5002H} capable of a log-scatter in measured $\dlgw$ of ${\sigma_{\ln D} = 0.05}$, and a high-z redshift survey with large sky coverage and negligible redshift uncertainties \citep[such as, e.g., SKA,][]{Bull:2018lat}.

The range of angular scales that we consider is limited by two factors. On small scales, large multipoles (${\ell>100}$) are excluded due to the angular resolution of about $1$ degree expected for our GW detector configuration of choice \citep{2019CQGra..36v5002H}. On large scales, we do not explore values of $\ell<10$ because our modelling does not take into account the relativistic effects dominating the signal at these scales. Nevertheless, we stress that these multipoles contribute relatively little information compared to larger multipoles since they are dominated by cosmic variance.

Our window functions are distributed in the redshift range $[0.1, 3]$. We assume $N_\mathrm{gal} = 12$ galaxy bins equally spaced in redshift, and $N_\mathrm{GW}=8$ GW luminosity distance bins equally spaced in $\dlgw$. We mention in particular that this choice is not completely arbitrary. The number of GW bins is motivated by forcing well-defined bins such that their width is at least three times the luminosity distance uncertainty $\sigma_{\ln D}$ that we have assumed. Furthermore, we have also verified that the exact number of galaxy bins does not dominate our results as long as $N_\mathrm{gal}>N_\mathrm{GW}$.

As for the last ingredient, we assume cosmic-variance or shot-noise limited uncertainties. In this case, we can write the covariance matrix of the auto-correlation and cross-correlation signals defined in Equations~\eqref{eq:autogw}, ~\eqref{eq:autog} and \eqref{eq:cc} as the following:
\begin{equation}
\begin{split}
    \mathrm{Cov} \left[ C^{ij}(\ell) C^{mn}(\ell^\prime) \right] &= \frac{\delta_{\ell \ell^\prime}}{(2\ell + 1)f_\mathrm{sky}} \times \\ 
     & \left( \tilde{C}^{im} \tilde{C}^{jn} + \tilde{C}^{in} \tilde{C}^{jm} \right),
     \label{eq:errors}
\end{split}
\end{equation}
where the indices $i, j, m, n$ can represent both galaxy or gravitational wave bins. The terms $\tilde{C}^{im}$ contain the shot-noise contribution when they represent the autocorrelations in the same bin:
\begin{equation}
    \tilde{C}^{im} (\ell) = C^{im} (\ell) + \frac{\delta_{im}}{\bar{n}},
\end{equation}
where $\bar{n}$ is the average density of projected objects from Equation~\eqref{eq:avgproj}. In this work, we  assume a survey covering a sky fraction equal to $f_\mathrm{sky} = 0.5$. 

Let us point out that we do not use the cross-bin correlations for the bins of the same type as a signal. However, we properly take into account the $\tilde{C}^{im}$ terms for overlaps between two different GW bins. Similar terms for galaxy bins are completely negligible as there are no overlaps between the spectroscopic redshift bins.

For the setup described in this section, we find a total SNR of the GW-gal and GW-GW angular power-spectra of $\sim 37$. This value is dominated by the GW-gal cross-correlations since the GW-GW auto-correlations are not well measured (SNR$\lesssim 6$).  

To generalize our choices, in \autoref{sec:SNR-scaling} we expand on how different combinations of instrumental specifications can affect the precision of the reconstruction.

\subsection{Reconstruction Pipeline}

In this section, we describe how the mock data presented in the previous section can be used to reconstruct $\dlgw$ and $\bgw$ as a function of $z$. For ease of interpretation and visualization, in our analysis we do not fit these functions directly, but instead focus on the ratios $\ratio\left(z\right)$ and $\bgw/b_\mathrm{gal}\left(z\right)$. We point out that we do not marginalize over different possibilities for $b_\mathrm{gal}\left(z\right)$. This is because we are not interested in exploring the properties of the galaxy population, which are expected to be very well constrained by the galaxy-galaxy correlation signal alone.

\begin{figure}[ht!]
    \centering
    \includegraphics[width=\columnwidth]{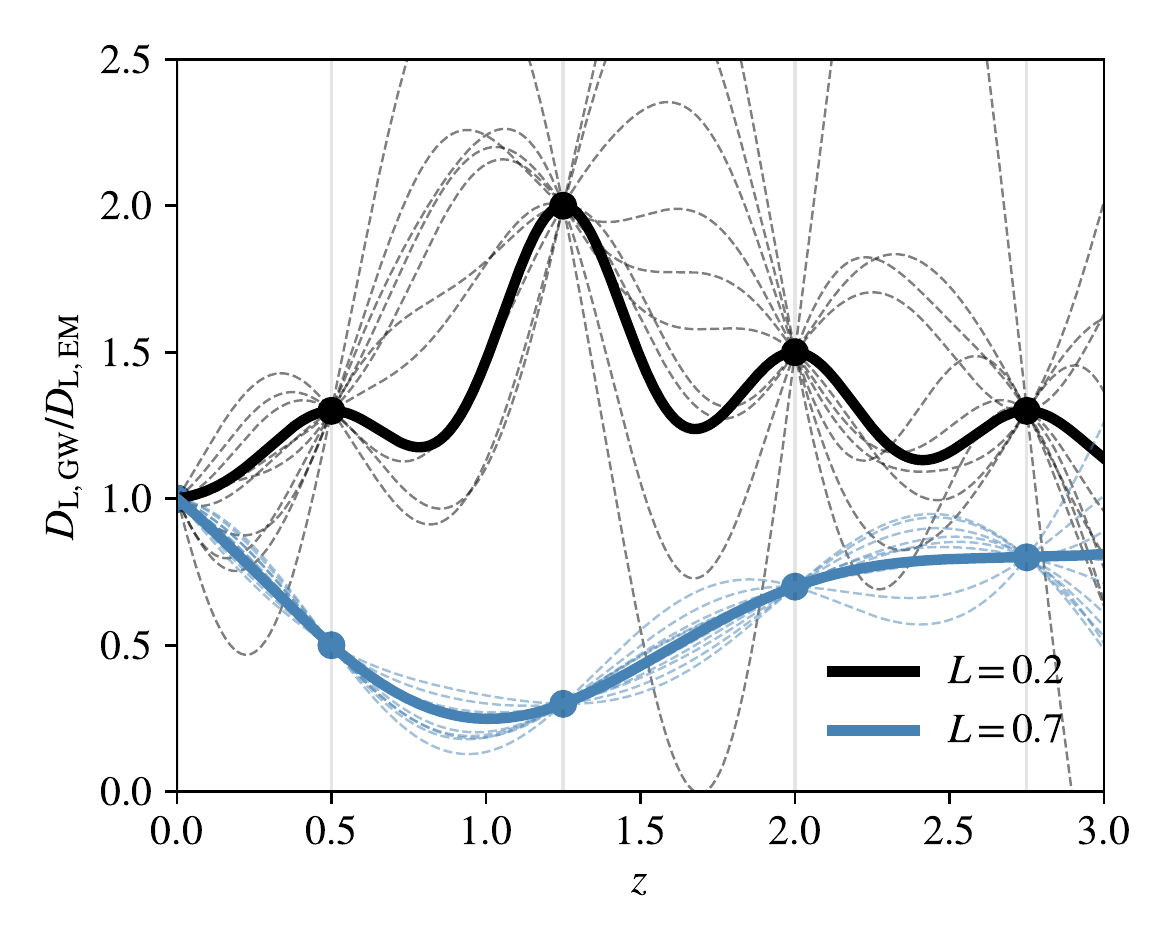}
    \caption{Example of GP reconstruction. The function $\ratio\left(z\right)$ is constructed using $4$ nodes at fixed redshifts (filled black dots). By varying the amplitudes of the nodes and the correlation length (L), it is possible to obtain different functional forms. The dashed lines represent a few possible realizations, while the thick lines represents the GP best fit from \texttt{sklearn} used in our sampling. Larger correlation lengths produce smoother lines.}
    \label{fig:reconstruction_example}
\end{figure}

As emphasized earlier, our approach makes use of a GP regression. This method is often used when the function of interest is directly measured at certain redshifts. These measurements are used as a training sample of the GP model, which then can predict the values of the reconstructed function at redshifts lacking any direct measurements \citep{Belgacem:2019zzu, renzi2020strongly, Mukherjee_2021, said2021reconstructing, perenon2021multitasking}.

However, this is not directly applicable for our current problem as neither the $\ratio\left(z\right)$ nor the $\bgw/b_\mathrm{gal}\left(z\right)$ are directly observable. Instead, these functions determine the auto- and cross-angular power spectra, which constitute our direct observables. In order to use GPs for our problem, we consider a certain number of redshift nodes for the two functions, referred to as \textit{training nodes} with a slight abuse of terminology. The amplitudes of the nodes are free and, given a node configuration, we consider GPs which pass through all of these nodes exactly. To render our scenario computationally feasible and not consider many functions for each node configuration, we use the GPs regressor of the \texttt{python} package \texttt{sklearn} to output the best fit and use this as our function.

Our use of GPs can be thought of as a binning of the functions of interest in redshift space, and imposing a certain prior correlations between the bins. These correlations are specified by the GP kernel function, which in our case is chosen to be 
\begin{equation}\label{eq:kernel}
    \kappa(z_i, z_{j};\,L) \propto \exp\left\{-\frac{1}{2}\left(\frac{\vert z_i - z_{j}\vert}{L}\right)^2\right\} ,
\end{equation}
where $L$ is the so-called correlation length. This kernel is flexible enough for our purposes, and we do not expect the detailed choice to have any significant impact on our results. For computational purposes, we generate the GPs using a baseline around $\ratio\left(z\right) = 1$. This baseline makes the GPs reconstruction to efficiently return to $\ratio\left(z\right) = 1$ when not pushed towards other values by the training nodes.

This process is described in \autoref{fig:reconstruction_example} for two values of correlation length $L$. While pictured in this example, for physical reasons in our analysis we exclude negative-valued functions when exploring $\ratio\left(z\right)$ and $\bgw/\bg\left(z\right)$, and also non-monotonic realizations of $\dlgw\left(z\right)$.

The goal of our statistical analysis is to explore the possible constraints on the shape of both $\ratio\left(z\right)$ and $\bgw/b_\mathrm{gal}\left(z\right)$. For that, we aim to sample the posterior distributions of the amplitudes of the nodes as well as the cosmological parameters. The correlation length $L$ can in principle be fixed based on theoretical priors. Lacking such priors in our case, we only impose a wide uniform prior on $L$ and consider it as a free parameter (see \autoref{tab:fiducial}).

Each step of the sampling process consists of producing two GP curves -- one for each $\ratio\left(z\right)$ and $\bgw/b_\mathrm{gal}\left(z\right)$ -- given the current set of training node amplitudes. The curves are used in the calculation of theoretical auto- and cross-correlation power spectra described in \autoref{sec:Cells}. The theoretical power spectra enter the Gaussian likelihood together with the generated mock data. The theoretical predictions are computed using our \texttt{python} code which is interfaced with the \texttt{emcee} sampler. The typical runs with varying cosmology take approximately 10 hours on a modern machine.

After obtaining the posterior distribution of the nodes, we reverse-engineer the problem to obtain confidence contours for each $\ratio\left(z\right)$ and $\bgw/b_\mathrm{gal}\left(z\right)$. For all the sampled node amplitudes we generate the corresponding GP profiles on finite but sufficiently many redshift points and calculate the $68\%$ and $95\%$ confidence intervals at each redshift using the statistical \texttt{python} package \texttt{GetDist}.

\begin{table}
    \centering
    \begin{ruledtabular}
    \begin{tabular}{c|c}
        Parameter & Prior 
        \\ \hline\hline
        Node amplitudes & [0, 11] (Uniform) \\ \hline
        Correlation length (L) & [1, 10] (Uniform) \\ \hline
        $\Omega_m$ & $1 \%$ (Gaussian) \\ \hline
        $h$ & $1 \%$ (Gaussian) \\ \hline
        $w_0$ & $5 \%$ (Gaussian)
    \end{tabular}
    \end{ruledtabular}
    \caption{Summary of the priors imposed before reconstructing $\bgw/\bg\left(z\right)$ and $\ratio\left(z\right)$ using $4$ nodes each. The GP hyper-parameters (i.e., the $2$ correlation lengths and the $4\times2$ amplitudes) are explored independently. The fiducial model is given by ${\Xi_0 = 1.4}, {n = 1}, {\Omega_\mathrm{m} = 0.31}, {h = 0.67}, {w_0 = -1}$.}
    \label{tab:fiducial}
\end{table}

To assess the impact of different cosmological backgrounds and clustering properties, we include in our reconstruction three nuisance parameters: the dark matter abundance $\Omega_\mathrm{m}$, the Hubble constant $H_0$ and the dark energy equation of state parameter $w_0$. A summary of our model parameters and priors used in this work is presented in \autoref{tab:fiducial}.

We would like to point out that upcoming galaxy surveys will measure these parameters with very high precision. While GWs alone might be able to provide competitive constraints, the focus of our analysis is not in constraining them. Rather, we would like to quantify how accurately the GW luminosity distance and source properties can be measured. These properties are not accessible to generic redshift surveys and can only be measured with the use of GW-specific observables. This reasoning justifies our tight Gaussian priors on the aforementioned cosmological parameters.

\section{Results}
\label{sec:results}
\subsection{Reconstructions}
\begin{figure}
    \centering
    \includegraphics[width=\columnwidth]{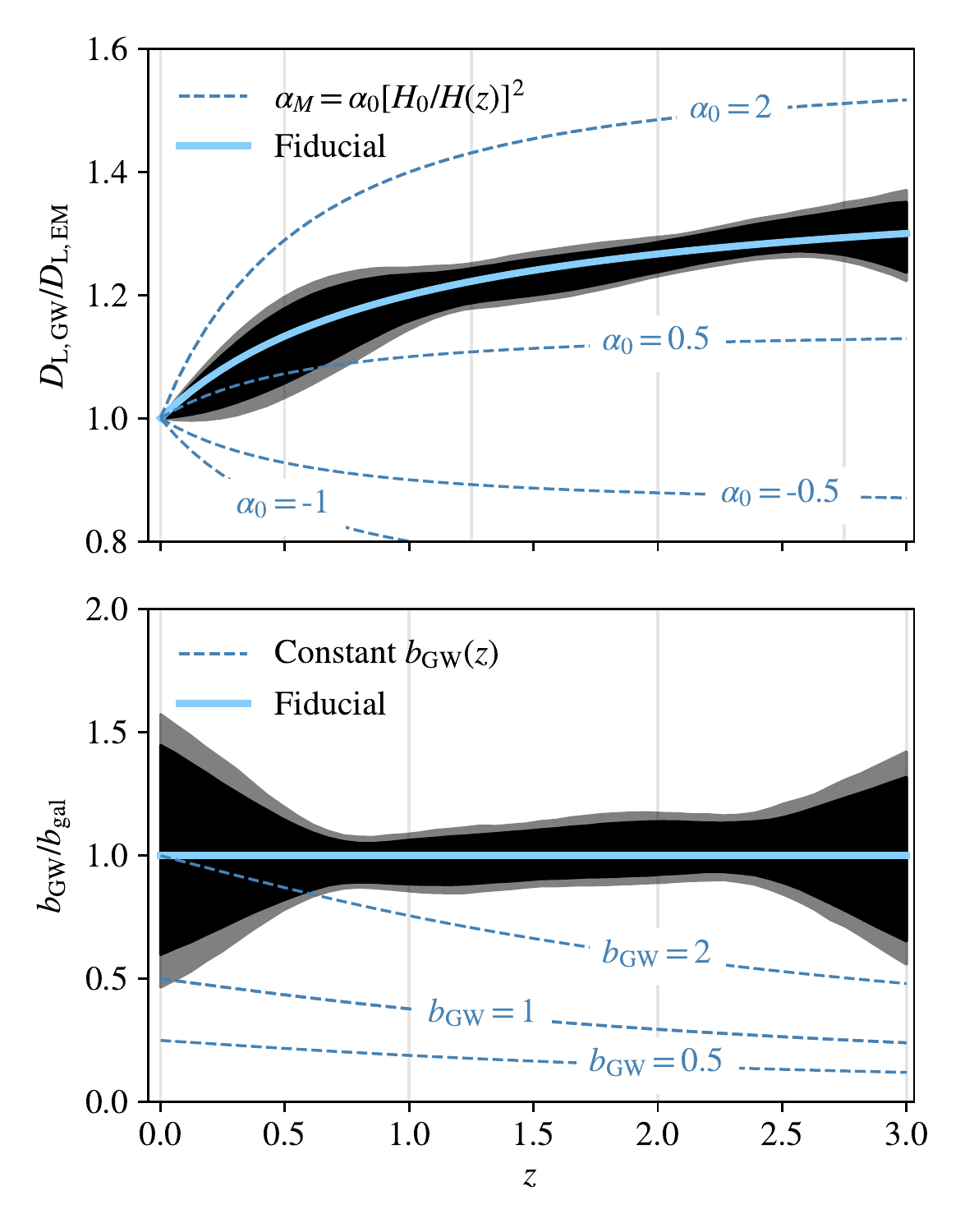}
    \caption{Confidence intervals ($68 \%$, in black, and $95 \%$, in lighter grey) of the jointly reconstructed functions $\ratio\left(z\right)$ and $\bgw/\bg\left(z\right)$. Together with the assumed fiducial model, we also plot the expectation for different models (see text for more details). The vertical lines mark the fixed location of the nodes used in the GP reconstruction.}
    \label{fig:reconstruction}
\end{figure}

In our reconstructions, we always impose the $\ratio\left(z\right)$ to become unity at redshift zero by placing a fixed node at $z = 0$ with an amplitude of $1$. Besides this fixed node, we have $4$ nodes for each of the reconstructed functions. We have arrived at this number by gradually increasing the number of nodes and monitoring the goodness of fit. In practice, we have monitored the AIC information criterion for $1$-, $2$- and $3$- node setups for the GW luminosity distance. Our experiments suggest that, as expected, the $1$-node configuration is significantly worse than the presented $4$-node setup. The $2$- and $3$- node setups have similar performance compared to the $4$-node case, but the latter still outperforms the former two. We then use the same number of nodes for the bias reconstruction. 

In our main analysis, the redshift positions $z_i$ of the nodes are fixed. We have, however, performed an experimental run to assess the impact of letting them free in reduced uniform prior ranges. The result of this experimentation is that the node locations remain unconstrained, and the final posterior of e.g. $\ratio\left(z\right)$ does not change when the node redshift locations are being sampled as free parameters. This implies that the exact locations of the GP notes are unimportant given they are uniformly distributed in the redshift range of interest.

It is worth emphasizing that when applying our methodology to real data, the number of nodes, as well as their exact redshift placements, should be constrained by performing similar, but more systematic experiments. Particularly, more accurate measures, such as the Bayesian evidence ratios, should be employed. Also, if enough data is used, some possible constraints could be found by letting the training nodes be completely free in the entire redshift range of interest. As the resulting posteriors are expected to be multimodal, this should be investigated using \textit{nested sampling} algorithms.       

As mentioned earlier, we also explore the GP correlation lengths both for the bias and the luminosity distance. On physical grounds, we are interested in smooth GP functions and have imposed the minimum of the uniform prior range for $L$ to be of the order of the inter-node distance so that the smoothness is maintained. We find that, as expected, both of the correlation lengths remain unconstrained within the imposed wide prior ranges.

The results of our joint reconstruction is presented in \autoref{fig:reconstruction}. In the same Figure, we also compare these constraints to different theoretical models. In the case of $\ratio\left(z\right)$, we use the parametrization
\begin{equation}
    \alpha_M(z) = \alpha_{0} \left[\frac{H_0}{H(z)}\right]^2,
\end{equation}
where we use the Equation~\eqref{eq:hubble} with $w_0 = -1$ to obtain the plotted lines \citep{Belgacem:2019pkk}. On the other hand, for $\bgw/\bg\left(z\right)$ we plot the lines corresponding to constant values of $b_\mathrm{GW}\left(z\right)$, while keeping the galaxy bias fixed to the expression in Equation~\eqref{eq:bg}. 

As expected, we observe how the fiducial models for both $\ratio\left(z\right)$ and $\ratiobias\left(z\right)$ are well encoded within the reconstructed confidence contours in both panels of \autoref{fig:reconstruction}. The constraints at higher redshift ($z\approx3$) for both reconstructions are broader. This is an effect that could not be seen if a parametric function was used for $\ratio\left(z\right)$, for instance, as the parametrization would have fixed the behaviour similarly at low and higher redshifts.

\begin{figure}
    \centering
    \includegraphics[width=\columnwidth]{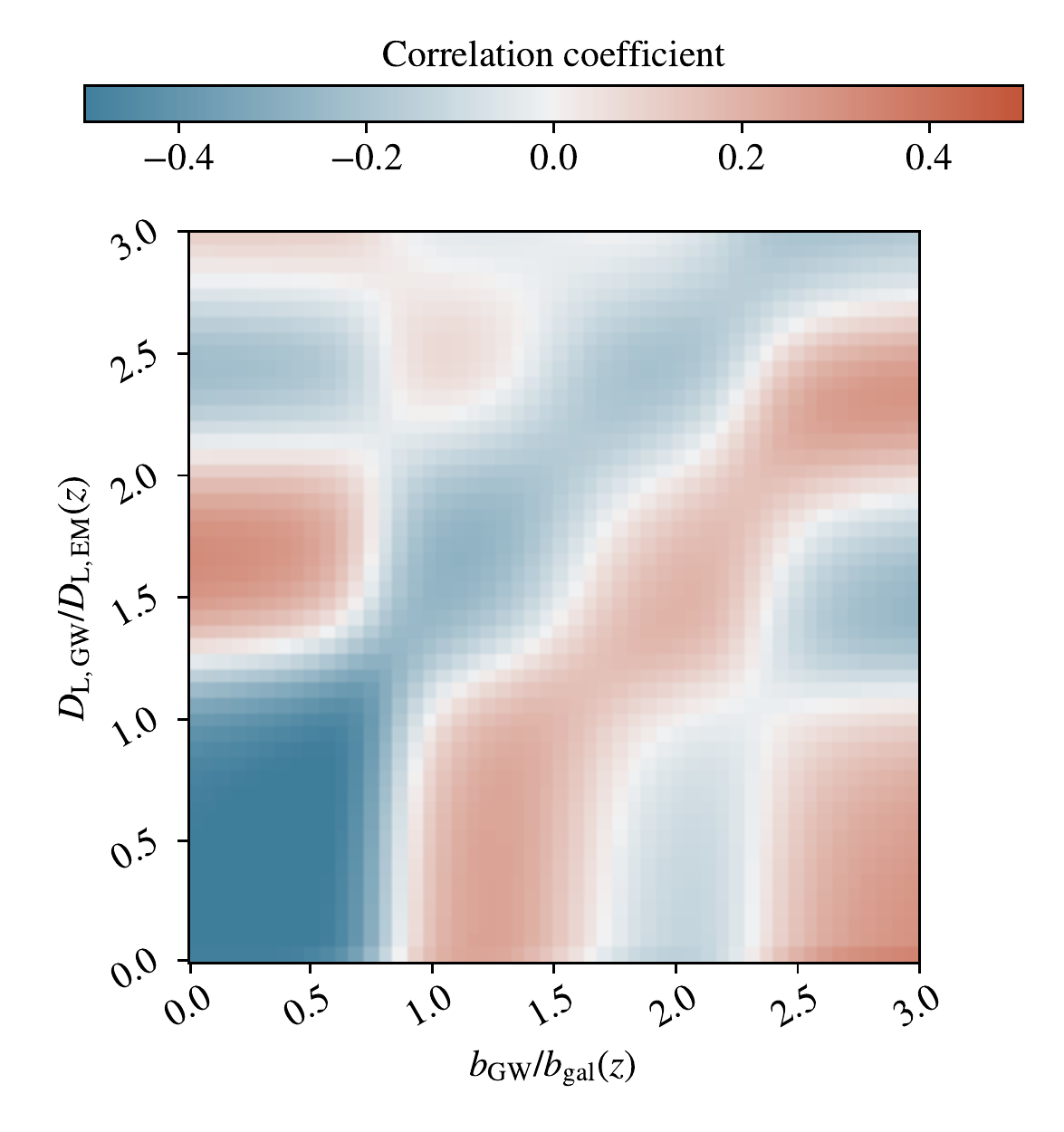}
    \caption{Correlation matrix for the reconstructed functions $\bgw/b_\mathrm{gal}\left(z\right)$ and $\ratio\left(z\right)$. The figure displays how mild, but non trivial correlations can arise from a joint reconstruction.}
    \label{fig:correlation}
\end{figure}

Finally, in \autoref{fig:correlation} we show the correlation between these functions. We observe weak but non-zero correlations between the GW bias and the luminosity distance. In general, parametric models might induce non-physical correlations. GPs are expected to behave better in this regard, but they can still induce spurious correlations due to finite correlation lengths. For consistency, we have also performed a reconstruction using a redshift binning approach and concluded that we can recover a very similar correlation structure with both of the methods.

\subsection{Signal-to-Noise scaling}
\label{sec:SNR-scaling}

The constraining power of our method crucially depends on a number of observational specifications. The most relevant parameters are 1) the angular sensitivity, specified by the maximum multipole $\ell_\mathrm{max}$ of the angular power spectra; 2) the number of GW sources, which is specified by the comoving number density $n_\mathrm{GW}$; and 3) the precision of the GW luminosity distance measurements $\sigma_{\ln D}$. In the case of $n_\mathrm{GW}$, we adjust the value of $n_0$ in Equation~\eqref{eq:ngw} as a way to explore different values of the total number of observed GW events, $N=4\pi f_\mathrm{sky}\bar{n}_\mathrm{GW}$. This, in principle, should include selection effects not captured by our formalism. Obviously, for a given experimental configuration the mentioned three variables are not independent, but it is still interesting to find the dependence of our results on each one of them separately. This allows us to reach conclusions without relying on specific experiments, and to suggest potential design guidelines for future GW detectors.

\begin{figure}
    \centering
    \includegraphics[width=\columnwidth]{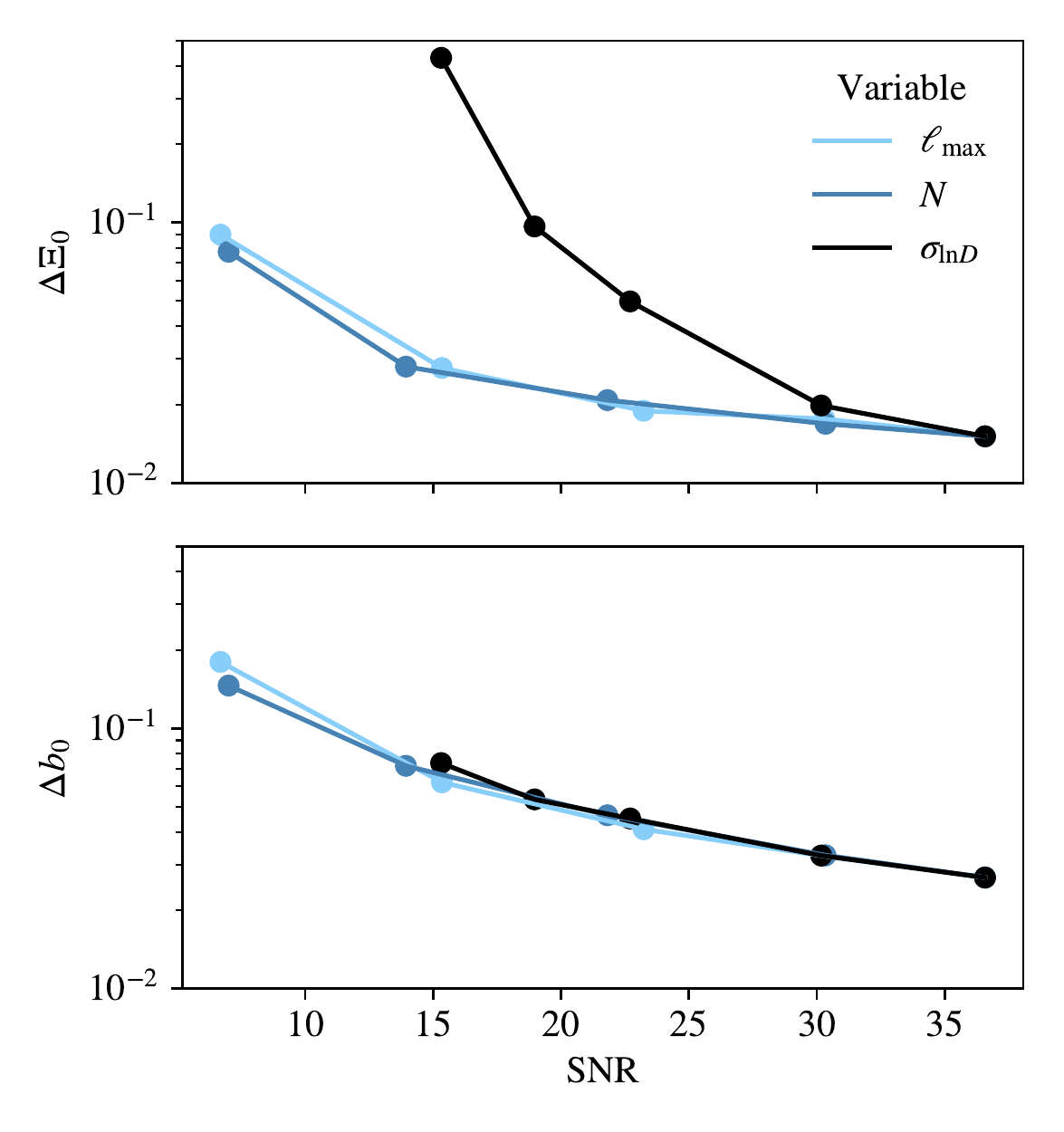}
    \caption{Scaling of the observed constraints with the cross-correlation SNR. Using a parametrized model for $\bgw/b_\mathrm{gal}\left(z\right)$ and $\ratio\left(z\right)$ we explore the constraining power of our method as a function of the number of observed GW sources $N = [0.7,  4,  7, 13, 20]\times 10^4$, angular resolution $\ell_\mathrm{max} = [20, 40, 60, 80, 100]$ and luminosity distance uncertainty $\sigma_{\ln D} = [0.5, 0.3, 0.2, 0.075, 0.05]$.  As visible from the figure, the data SNR completely captures  the effect on the observed uncertainties $\Delta b_0$ and $\Delta \Xi_0$ in the first two cases. In the case of $\sigma_{\ln D}$, we observe that the increase in constraining power for $\ratio\left(z\right)$ is steeper due to the larger number of window functions that we can build to sample $\dlgw\left(z\right)$.}
    \label{fig:SNRscaling}
\end{figure}

To attain such insights, in this subsection we consider constraints on the parametric expression in Equation~\eqref{eq:dlratioxi}, as well as the parametric GW bias given by Equation~\eqref{eq:biasgw}. For simplicity, we fix $n = 1$ and only constrain the parameter $\Xi_0$.

When varying $\ell_\mathrm{max}$ and $N$ we keep the rest of the configuration (including the luminosity distance binning) fixed. Each case of $\sigma_{\ln D}$, on the other hand, is accompanied by an adjustment in the number of luminosity distance bins. This is done to be consistent with our binning strategy, namely that the luminosity distance width of each bin is at least $\mathcal{O}(3)$ times wider than $\sigma_{\ln D}$.     

Our results are summarized in \autoref{fig:SNRscaling}, where we plot the anticipated uncertainties in $\Xi_0$ (upper panel) and $b_0$ (lower panel) as a function of the SNR of the cross-correlation in Equation~\eqref{eq:cc}.

For a fixed $\sigma_{\ln D}$ the constraining power on $\Xi_0$ and $b_0$ is almost completely determined by the cross-correlation SNR. This fact suggests that no matter how the given SNR is realized (either by increasing the number of sources or by improving the angular sensitivity), the expected constraints will be the same. This implies that the results presented in this paper can be easily scaled to different configurations. Unsurprisingly, we find that the constraints scale as $1/$SNR.

The situation is somewhat different for the case of varying $\sigma_{\ln D}$ (and the number of luminosity distance bins). The constraints on the bias still follow the same form (see the lower panel), but the scaling of the $\Xi_0$ constraints, on the other hand, is much steeper than in the cases of varying $\ell_\mathrm{max}$ and $n_\mathrm{GW}$, roughly $1/\mathrm{SNR}^3$. This fact can be qualitatively understood by remembering the importance of the relative positions of GW and galaxy window functions demonstrated in \autoref{fig:cross_corr_demo}. Sampling this relation with a larger number of window functions increases the precision of our reconstruction. 

The results presented in this section quantify the importance of accurate luminosity distance measurements and demonstrate the benefit that smaller values of $\sigma_{\ln D}$ can bring to a binned approach.

\section{Discussion and Conclusions}
\label{sec:discussion}

In this paper, we have shown how the combination of GW observations and redshift surveys can be exploited in the era of GW cosmology. We have identified two essential quantities that characterize this new research field: 1) the luminosity distance relation, a clear imprint of modifications to the propagation of tensor modes (see \autoref{sec:GWs}), and 2) the bias of the sources, regulating their spatial distribution and betraying their origin. 

Proposed GW detectors such as the Einstein Telescope \citep{Maggiore:2019uih} or Cosmic Explorer \citep{Reitze:2019iox} are expected to probe a sizeable fraction of the visible Universe and produce large statistical samples. In this context, we point out that the number of sources assumed for our main analysis, $2\times10^5$, is particularly conservative and differs from expectations by at least an order of magnitude \citep{Maggiore:2019uih}. This difference is primarily due to our assumption of a constant comoving density of events. While we do not explore other assumptions for the distribution of GW sources over cosmic time, we have investigated similar effects in \autoref{sec:SNR-scaling}, where we have shown how our results can be rescaled to other instrument configurations or number of observed sources. 

Our formalism, based on binned angular power-spectra and sky maps, is optimal for a large number of sources with no known counterpart. 
Its main advantages are related to the simple modelling of the theoretical signals and their data covariance matrix. Because no reconstruction of the underlying density field is necessary, the predictions display a clear separation of scales. For example, the angular scales that we have considered hare are all within the linear regime ($k\lesssim 0.1$ Mpc$^{-1}$). Furthermore, because this formalism is well established, our shot-noise limited covariance matrix can be easily generalized to include additional sources of (co-)variance.

Although a comprehensive comparison between multiple approaches is outside the scope of this work, it is worth discussing how our results compare to others found in the literature. We preface this by saying that one-to-one comparisons, however, are often complicated either by significantly different assumptions or the impossibility of directly translating these assumptions from one prescription to another. Despite this, here we draw a parallel between our method and two other methods.

The first method is the one used in \cite{Mukherjee:2020mha}, which has also been shown to be extremely successful in measuring both $\bgw\left(z\right)$ and $\dlgw\left(z\right)$ using parametric models. Similarly to this work, the information is also extracted from the cross-correlation with redshift sources, but no binning of the GW data is performed. In this case, we have verified that such methods perform significantly better than our map-based approach in the case of a low number density of GW sources and large uncertainty in the measured $\dlem\left(z\right)$. These features, in particular, make it especially useful for near-future samples of a few tens of objects. 

The second promising method to measure $\dlgw\left(z\right)$ that has been proposed in the literature is offered by GW sources with known counterparts. Such observations give direct access to $\dlgw$ as a function of redshift and can be combined with similar measurements in the EM spectrum to obtain $\ratio\left(z\right)$. The analysis of \cite{Belgacem:2019zzu} is based on this methodology and, similarly to ours, also employs GPs to reconstruct this ratio from an Einstein Telescope sample with $\sim 10^2$ sources. 

Ultimately, we expect this counterpart-based formalism and the one described in this work to be complementary: a direct measure of $\dlgw\left(z\right)$ can be used to break the degeneracy between bias and luminosity distance shown in \autoref{fig:correlation}. However, because the fraction of events with known counterparts that will be observed is heavily dependent on both the GW source distribution and multiple instrumental setups, we do not attempt to combine the two methods here. 

In conclusion, the combination of GW resolved events and the clustering of galaxies is expected to improve our current knowledge of the physical properties of the Universe. Our work shows how to reconstruct these properties as a function of redshift in a generic way, and highlights the need for accurate and precise measurements of $\dlgw$. This will require control over the instrument calibration uncertainties \citep{2017PhRvD..96j2001C}, but also the degeneracy between the inclination of the source and its luminosity distance  \citep{Ghosh:2015jra}. In the future, we aim to apply our current analysis pipeline to the next generation of large scale structure surveys and incoming GW observations.

\begin{acknowledgments}
We thank Masamune Oguri for a useful conversation. GCH and OC are supported by a de Sitter fellowship from the  Dutch Research Council (NWO). GCH acknowledges support from the Delta Institute for Theoretical Physics (D-ITP consortium) a program by the NWO. The work of VV is financed by the WPI Research Center Initiative, MEXT, Japan and by JSPS KAKENHI grant number 20K22348.
\end{acknowledgments}

\bibliography{references}{}

\begin{thebibliography}{}
\expandafter\ifx\csname natexlab\endcsname\relax\def\natexlab#1{#1}\fi
\providecommand{\url}[1]{\href{#1}{#1}}
\providecommand{\dodoi}[1]{doi:~\href{http://doi.org/#1}{\nolinkurl{#1}}}
\providecommand{\doeprint}[1]{\href{http://ascl.net/#1}{\nolinkurl{http://ascl.net/#1}}}
\providecommand{\doarXiv}[1]{\href{https://arxiv.org/abs/#1}{\nolinkurl{https://arxiv.org/abs/#1}}}

\bibitem[{Abbott {et~al.}(2016)}]{Abbott:2016blz}
Abbott, B.~P., {et~al.} 2016, Phys. Rev. Lett., 116, 061102,
  \dodoi{10.1103/PhysRevLett.116.061102}

\bibitem[{Abbott {et~al.}(2017{\natexlab{a}})}]{TheLIGOScientific:2017qsa}
---. 2017{\natexlab{a}}, Phys. Rev. Lett., 119, 161101,
  \dodoi{10.1103/PhysRevLett.119.161101}

\bibitem[{Abbott {et~al.}(2017{\natexlab{b}})}]{Abbott:2017xzu}
---. 2017{\natexlab{b}}, Nature, 551, 85, \dodoi{10.1038/nature24471}

\bibitem[{Abbott {et~al.}(2021)}]{Abbott:2019yzh}
---. 2021, Astrophys. J., 909, 218, \dodoi{10.3847/1538-4357/abdcb7}

\bibitem[{Abbott {et~al.}(2020)}]{Abbott:2020gyp}
Abbott, R., {et~al.} 2020.
\newblock \doarXiv{2010.14533}

\bibitem[{Aghanim {et~al.}(2020)}]{Akrami:2018vks}
Aghanim, N., {et~al.} 2020, Astron. Astrophys., 641, A1,
  \dodoi{10.1051/0004-6361/201833880}

\bibitem[{Akrami {et~al.}(2018)Akrami, Brax, Davis, \&
  Vardanyan}]{Akrami:2018yjz}
Akrami, Y., Brax, P., Davis, A.-C., \& Vardanyan, V. 2018, Phys. Rev. D, 97,
  124010, \dodoi{10.1103/PhysRevD.97.124010}

\bibitem[{Alonso {et~al.}(2020)Alonso, Cusin, Ferreira, \&
  Pitrou}]{Alonso:2020mva}
Alonso, D., Cusin, G., Ferreira, P.~G., \& Pitrou, C. 2020, Phys. Rev. D, 102,
  023002, \dodoi{10.1103/PhysRevD.102.023002}

\bibitem[{Amendola {et~al.}(2018)Amendola, Sawicki, Kunz, \&
  Saltas}]{Amendola_2018}
Amendola, L., Sawicki, I., Kunz, M., \& Saltas, I.~D. 2018, Journal of
  Cosmology and Astroparticle Physics, 2018, 030–030,
  \dodoi{10.1088/1475-7516/2018/08/030}

\bibitem[{Baker {et~al.}(2017)Baker, Bellini, Ferreira, Lagos, Noller, \&
  Sawicki}]{Baker_2017}
Baker, T., Bellini, E., Ferreira, P., {et~al.} 2017, Physical Review Letters,
  119, \dodoi{10.1103/physrevlett.119.251301}

\bibitem[{Baker \& Harrison(2021)}]{Baker:2020apq}
Baker, T., \& Harrison, I. 2021, JCAP, 01, 068,
  \dodoi{10.1088/1475-7516/2021/01/068}

\bibitem[{Belgacem {et~al.}(2018)Belgacem, Dirian, Foffa, \&
  Maggiore}]{Belgacem:2018lbp}
Belgacem, E., Dirian, Y., Foffa, S., \& Maggiore, M. 2018, Phys. Rev. D, 98,
  023510, \dodoi{10.1103/PhysRevD.98.023510}

\bibitem[{Belgacem {et~al.}(2020)Belgacem, Foffa, Maggiore, \&
  Yang}]{Belgacem:2019zzu}
Belgacem, E., Foffa, S., Maggiore, M., \& Yang, T. 2020, Phys. Rev. D, 101,
  063505, \dodoi{10.1103/PhysRevD.101.063505}

\bibitem[{Belgacem {et~al.}(2019)}]{Belgacem:2019pkk}
Belgacem, E., {et~al.} 2019, JCAP, 07, 024,
  \dodoi{10.1088/1475-7516/2019/07/024}

\bibitem[{Bellini \& Sawicki(2014)}]{Bellini:2014fua}
Bellini, E., \& Sawicki, I. 2014, JCAP, 07, 050,
  \dodoi{10.1088/1475-7516/2014/07/050}

\bibitem[{Bera {et~al.}(2020)Bera, Rana, More, \& Bose}]{Bera:2020jhx}
Bera, S., Rana, D., More, S., \& Bose, S. 2020, Astrophys. J., 902, 79,
  \dodoi{10.3847/1538-4357/abb4e0}

\bibitem[{Bird {et~al.}(2016)Bird, Cholis, Mu\~noz, Ali-Ha\"\i{}moud,
  Kamionkowski, Kovetz, Raccanelli, \& Riess}]{Bird:2016dcv}
Bird, S., Cholis, I., Mu\~noz, J.~B., {et~al.} 2016, Phys. Rev. Lett., 116,
  201301, \dodoi{10.1103/PhysRevLett.116.201301}

\bibitem[{Bustillo {et~al.}(2021)Bustillo, Sanchis-Gual, Torres-Forn\'e, Font,
  Vajpeyi, Smith, Herdeiro, Radu, \& Leong}]{CalderonBustillo:2020srq}
Bustillo, J.~C., Sanchis-Gual, N., Torres-Forn\'e, A., {et~al.} 2021, Phys.
  Rev. Lett., 126, 081101, \dodoi{10.1103/PhysRevLett.126.081101}

\bibitem[{Ca\~nas Herrera {et~al.}(2020{\natexlab{a}})Ca\~nas Herrera,
  Contigiani, \& Vardanyan}]{Canas-Herrera:2019npr}
Ca\~nas Herrera, G., Contigiani, O., \& Vardanyan, V. 2020{\natexlab{a}}, Phys.
  Rev. D, 102, 043513, \dodoi{10.1103/PhysRevD.102.043513}

\bibitem[{Ca\~nas Herrera {et~al.}(2020{\natexlab{b}})Ca\~nas Herrera, Torrado,
  \& Ach\'ucarro}]{Canas-Herrera:2020mme}
Ca\~nas Herrera, G., Torrado, J., \& Ach\'ucarro, A. 2020{\natexlab{b}}.
\newblock \doarXiv{2012.04640}

\bibitem[{{Cahillane} {et~al.}(2017){Cahillane}, {Betzwieser}, {Brown},
  {Goetz}, {Hall}, {Izumi}, {Kandhasamy}, {Karki}, {Kissel}, {Mendell},
  {Savage}, {Tuyenbayev}, {Urban}, {Viets}, {Wade}, \&
  {Weinstein}}]{2017PhRvD..96j2001C}
{Cahillane}, C., {Betzwieser}, J., {Brown}, D.~A., {et~al.} 2017, \prd, 96,
  102001, \dodoi{10.1103/PhysRevD.96.102001}

\bibitem[{Calcagni {et~al.}(2019)Calcagni, Kuroyanagi, Marsat, Sakellariadou,
  Tamanini, \& Tasinato}]{Calcagni:2019ngc}
Calcagni, G., Kuroyanagi, S., Marsat, S., {et~al.} 2019, JCAP, 10, 012,
  \dodoi{10.1088/1475-7516/2019/10/012}

\bibitem[{Clesse \& Garc\'\i{}a-Bellido(2017)}]{Clesse:2016vqa}
Clesse, S., \& Garc\'\i{}a-Bellido, J. 2017, Phys. Dark Univ., 15, 142,
  \dodoi{10.1016/j.dark.2016.10.002}

\bibitem[{Contigiani(2020)}]{Contigiani:2020yyc}
Contigiani, O. 2020, Mon. Not. Roy. Astron. Soc., 492, 3359,
  \dodoi{10.1093/mnras/staa026}

\bibitem[{Creminelli \& Vernizzi(2017)}]{Creminelli:2017sry}
Creminelli, P., \& Vernizzi, F. 2017, Phys. Rev. Lett., 119, 251302,
  \dodoi{10.1103/PhysRevLett.119.251302}

\bibitem[{Crittenden {et~al.}(2009)Crittenden, Pogosian, \&
  Zhao}]{Crittenden_2009}
Crittenden, R.~G., Pogosian, L., \& Zhao, G.-B. 2009, Journal of Cosmology and
  Astroparticle Physics, 2009, 025–025, \dodoi{10.1088/1475-7516/2009/12/025}

\bibitem[{Crittenden {et~al.}(2012)Crittenden, Zhao, Pogosian, Samushia, \&
  Zhang}]{Crittenden_2012}
Crittenden, R.~G., Zhao, G.-B., Pogosian, L., Samushia, L., \& Zhang, X. 2012,
  Journal of Cosmology and Astroparticle Physics, 2012, 048–048,
  \dodoi{10.1088/1475-7516/2012/02/048}

\bibitem[{Deffayet \& Menou(2007)}]{Deffayet:2007kf}
Deffayet, C., \& Menou, K. 2007, Astrophys. J. Lett., 668, L143,
  \dodoi{10.1086/522931}

\bibitem[{Diemer(2018)}]{Diemer:2017bwl}
Diemer, B. 2018, Astrophys. J. Suppl., 239, 35,
  \dodoi{10.3847/1538-4365/aaee8c}

\bibitem[{Dirian {et~al.}(2016)Dirian, Foffa, Kunz, Maggiore, \&
  Pettorino}]{Dirian_2016}
Dirian, Y., Foffa, S., Kunz, M., Maggiore, M., \& Pettorino, V. 2016, Journal
  of Cosmology and Astroparticle Physics, 2016, 068–068,
  \dodoi{10.1088/1475-7516/2016/05/068}

\bibitem[{Ezquiaga \& Zumalac\'arregui(2017)}]{Ezquiaga:2017ekz}
Ezquiaga, J.~M., \& Zumalac\'arregui, M. 2017, Phys. Rev. Lett., 119, 251304,
  \dodoi{10.1103/PhysRevLett.119.251304}

\bibitem[{Ezquiaga \& Zumalac\'arregui(2020)}]{Ezquiaga:2020dao}
---. 2020, Phys. Rev. D, 102, 124048, \dodoi{10.1103/PhysRevD.102.124048}

\bibitem[{Finke {et~al.}(2021)Finke, Foffa, Iacovelli, Maggiore, \&
  Mancarella}]{Finke:2021aom}
Finke, A., Foffa, S., Iacovelli, F., Maggiore, M., \& Mancarella, M. 2021.
\newblock \doarXiv{2101.12660}

\bibitem[{Foreman-Mackey {et~al.}(2013)Foreman-Mackey, Hogg, Lang, \&
  Goodman}]{ForemanMackey:2012ig}
Foreman-Mackey, D., Hogg, D.~W., Lang, D., \& Goodman, J. 2013, Publ. Astron.
  Soc. Pac., 125, 306, \dodoi{10.1086/670067}

\bibitem[{Garoffolo {et~al.}(2020)Garoffolo, Tasinato, Carbone, Bertacca, \&
  Matarrese}]{Garoffolo:2019mna}
Garoffolo, A., Tasinato, G., Carbone, C., Bertacca, D., \& Matarrese, S. 2020,
  JCAP, 11, 040, \dodoi{10.1088/1475-7516/2020/11/040}

\bibitem[{Gerardi {et~al.}(2019)Gerardi, Martinelli, \&
  Silvestri}]{Gerardi:2019obr}
Gerardi, F., Martinelli, M., \& Silvestri, A. 2019, JCAP, 07, 042,
  \dodoi{10.1088/1475-7516/2019/07/042}

\bibitem[{Ghosh {et~al.}(2016)Ghosh, Del~Pozzo, \& Ajith}]{Ghosh:2015jra}
Ghosh, A., Del~Pozzo, W., \& Ajith, P. 2016, Phys. Rev. D, 94, 104070,
  \dodoi{10.1103/PhysRevD.94.104070}

\bibitem[{{Hall} \& {Evans}(2019)}]{2019CQGra..36v5002H}
{Hall}, E.~D., \& {Evans}, M. 2019, Classical and Quantum Gravity, 36, 225002,
  \dodoi{10.1088/1361-6382/ab41d6}

\bibitem[{Holz \& Hughes(2005)}]{Holz:2005df}
Holz, D.~E., \& Hughes, S.~A. 2005, Astrophys. J., 629, 15,
  \dodoi{10.1086/431341}

\bibitem[{Horndeski(1974)}]{Horndeski:1974wa}
Horndeski, G.~W. 1974, Int. J. Theor. Phys., 10, 363,
  \dodoi{10.1007/BF01807638}

\bibitem[{Lewis(2019)}]{getdist}
Lewis, A. 2019.
\newblock \doarXiv{1910.13970}

\bibitem[{Li {et~al.}(2021)Li, Wang, Han, Tang, Yuan, Fan, \& Wei}]{Li:2021ukd}
Li, Y.-J., Wang, Y.-Z., Han, M.-Z., {et~al.} 2021.
\newblock \doarXiv{2104.02969}

\bibitem[{Lombriser \& Taylor(2016)}]{Lombriser_2016}
Lombriser, L., \& Taylor, A. 2016, Journal of Cosmology and Astroparticle
  Physics, 2016, 031–031, \dodoi{10.1088/1475-7516/2016/03/031}

\bibitem[{Maggiore {et~al.}(2020)}]{Maggiore:2019uih}
Maggiore, M., {et~al.} 2020, JCAP, 03, 050,
  \dodoi{10.1088/1475-7516/2020/03/050}

\bibitem[{Max {et~al.}(2017)Max, Platscher, \& Smirnov}]{Max_2017}
Max, K., Platscher, M., \& Smirnov, J. 2017, Physical Review Letters, 119,
  \dodoi{10.1103/physrevlett.119.111101}

\bibitem[{Max {et~al.}(2018)Max, Platscher, \& Smirnov}]{Max_2018}
---. 2018, Physical Review D, 97, \dodoi{10.1103/physrevd.97.064009}

\bibitem[{Mukherjee \& Mukherjee(2021)}]{Mukherjee_2021}
Mukherjee, P., \& Mukherjee, A. 2021, Monthly Notices of the Royal Astronomical
  Society, \dodoi{10.1093/mnras/stab1054}

\bibitem[{Mukherjee \& Wandelt(2018)}]{Mukherjee:2018ebj}
Mukherjee, S., \& Wandelt, B.~D. 2018.
\newblock \doarXiv{1808.06615}

\bibitem[{Mukherjee {et~al.}(2020)Mukherjee, Wandelt, \&
  Silk}]{Mukherjee:2020mha}
Mukherjee, S., Wandelt, B.~D., \& Silk, J. 2020, \dodoi{10.1093/mnras/stab001}

\bibitem[{Oguri(2016)}]{Oguri:2016dgk}
Oguri, M. 2016, Phys. Rev. D, 93, 083511, \dodoi{10.1103/PhysRevD.93.083511}

\bibitem[{Perenon {et~al.}(2021)Perenon, Martinelli, Ili\'c, Maartens, Lochner,
  \& Clarkson}]{perenon2021multitasking}
Perenon, L., Martinelli, M., Ili\'c, S., {et~al.} 2021.
\newblock \doarXiv{2105.01613}

\bibitem[{Price-Whelan {et~al.}(2018)}]{Price-Whelan:2018hus}
Price-Whelan, A.~M., {et~al.} 2018, Astron. J., 156, 123,
  \dodoi{10.3847/1538-3881/aabc4f}

\bibitem[{Raccanelli {et~al.}(2016)Raccanelli, Kovetz, Bird, Cholis, \&
  Munoz}]{Raccanelli:2016cud}
Raccanelli, A., Kovetz, E.~D., Bird, S., Cholis, I., \& Munoz, J.~B. 2016,
  Phys. Rev. D, 94, 023516, \dodoi{10.1103/PhysRevD.94.023516}

\bibitem[{Raccanelli {et~al.}(2018)Raccanelli, Vidotto, \&
  Verde}]{Raccanelli:2017xee}
Raccanelli, A., Vidotto, F., \& Verde, L. 2018, JCAP, 08, 003,
  \dodoi{10.1088/1475-7516/2018/08/003}

\bibitem[{Rasmussen \& Williams(2005)}]{ML}
Rasmussen, C.~E., \& Williams, C. K.~I. 2005, Gaussian Processes for Machine
  Learning (Adaptive Computation and Machine Learning) (The MIT Press)

\bibitem[{Reitze {et~al.}(2019)}]{Reitze:2019iox}
Reitze, D., {et~al.} 2019, Bull. Am. Astron. Soc., 51, 035.
\newblock \doarXiv{1907.04833}

\bibitem[{Renzi {et~al.}(2020)Renzi, Hogg, Martinelli, \&
  Nesseris}]{renzi2020strongly}
Renzi, F., Hogg, N.~B., Martinelli, M., \& Nesseris, S. 2020, Strongly lensed
  supernovae as a self-sufficient probe of the distance duality relation.
\newblock \doarXiv{2010.04155}

\bibitem[{Robitaille {et~al.}(2013)}]{Robitaille:2013mpa}
Robitaille, T.~P., {et~al.} 2013, Astron. Astrophys., 558, A33,
  \dodoi{10.1051/0004-6361/201322068}

\bibitem[{Said {et~al.}(2021)Said, Mifsud, Sultana, \&
  Adami}]{said2021reconstructing}
Said, J.~L., Mifsud, J., Sultana, J., \& Adami, K.~Z. 2021, Reconstructing
  teleparallel gravity with cosmic structure growth and expansion rate data.
\newblock \doarXiv{2103.05021}

\bibitem[{Sakstein \& Jain(2017)}]{Sakstein_2017}
Sakstein, J., \& Jain, B. 2017, Physical Review Letters, 119,
  \dodoi{10.1103/physrevlett.119.251303}

\bibitem[{Sasaki {et~al.}(2016)Sasaki, Suyama, Tanaka, \&
  Yokoyama}]{Sasaki:2016jop}
Sasaki, M., Suyama, T., Tanaka, T., \& Yokoyama, S. 2016, Phys. Rev. Lett.,
  117, 061101, \dodoi{10.1103/PhysRevLett.117.061101}

\bibitem[{Sasaki {et~al.}(2018)Sasaki, Suyama, Tanaka, \&
  Yokoyama}]{Sasaki_2018}
---. 2018, Classical and Quantum Gravity, 35, 063001,
  \dodoi{10.1088/1361-6382/aaa7b4}

\bibitem[{Scelfo {et~al.}(2018)Scelfo, Bellomo, Raccanelli, Matarrese, \&
  Verde}]{Scelfo:2018sny}
Scelfo, G., Bellomo, N., Raccanelli, A., Matarrese, S., \& Verde, L. 2018,
  JCAP, 09, 039, \dodoi{10.1088/1475-7516/2018/09/039}

\bibitem[{Scelfo {et~al.}(2020)Scelfo, Boco, Lapi, \& Viel}]{Scelfo:2020jyw}
Scelfo, G., Boco, L., Lapi, A., \& Viel, M. 2020, JCAP, 10, 045,
  \dodoi{10.1088/1475-7516/2020/10/045}

\bibitem[{Schutz(1986)}]{Schutz:1986gp}
Schutz, B.~F. 1986, Nature, 323, 310, \dodoi{10.1038/323310a0}

\bibitem[{Scikit-learn(2018)}]{sklearn}
Scikit-learn. 2018, {Scikit-learn 0.19.1 documentation: Gaussian Processes},
  \url{http://scikit-learn.org/stable/modules/gaussian_process.html}

\bibitem[{Shafieloo {et~al.}(2012)Shafieloo, Kim, \& Linder}]{Shafieloo_2012}
Shafieloo, A., Kim, A.~G., \& Linder, E.~V. 2012, Physical Review D, 85,
  \dodoi{10.1103/physrevd.85.123530}

\bibitem[{Soares-Santos {et~al.}(2019)}]{Soares-Santos:2019irc}
Soares-Santos, M., {et~al.} 2019, Astrophys. J. Lett., 876, L7,
  \dodoi{10.3847/2041-8213/ab14f1}

\bibitem[{Weltman {et~al.}(2020)}]{Bull:2018lat}
Weltman, A., {et~al.} 2020, Publ. Astron. Soc. Austral., 37, e002,
  \dodoi{10.1017/pasa.2019.42}

\bibitem[{Zhao {et~al.}(2012)Zhao, Crittenden, Pogosian, \& Zhang}]{Zhao_2012}
Zhao, G.-B., Crittenden, R.~G., Pogosian, L., \& Zhang, X. 2012, Physical
  Review Letters, 109, \dodoi{10.1103/physrevlett.109.171301}

\end{thebibliography}
\bibliographystyle{aasjournal}

\end{document}